\documentclass[pra,twocolumn,showpacs,preprintnumbers,
amsmath,amssymb,floatfix]{revtex4}
\usepackage{hyperref}
\usepackage{amsmath}
\usepackage[dvips]{epsfig}

\newcommand{\beq}{\begin{equation}}
\newcommand{\eeq}{\end{equation}} 
\newcommand{\beqa}{\begin{eqnarray}}
\newcommand{\eeqa}{\end{eqnarray}}
\newcommand{\ba}{\begin{array}}
\newcommand{\ea}{\end{array}}

\begin{document}
\title{Supersolid structure and excitation spectrum of soft-core bosons in 3D}
\author{Francesco Ancilotto$^{1,2}$, Maurizio Rossi$^{1}$ and Flavio Toigo$^{1,2}$} 
\affiliation{$^1$Dipartimento di Fisica e Astronomia 
"Galileo Galilei" and CNISM, Universit\`a di Padova, 
Via Marzolo 8, 35122 Padova, Italy \\
$^2$CNR-IOM Democritos, via Bonomea, 265 - 34136 Trieste, Italy} 

\begin{abstract} 
By means of a mean-field method, we have studied the 
zero temperature structure and excitation spectrum 
of a three-dimensional soft-core bosonic system for 
a value of the interaction strength 
that favors a crystal structure made of atomic nano-clusters  
arranged with FCC ordering. In addition 
to the longitudinal and transverse 
phonon branches expected for a normal crystal, the excitation spectrum shows 
a soft mode related to the breaking of gauge symmetry, 
which signals a partial superfluid character of the solid.
Additional evidence of supersolidity is provided by the
calculation of the superfluid fraction, which shows
a first-order drop, from 1 to 0.4, at the liquid-supersolid 
transition and a monotonic 
decrease as the interaction strength parameter is increased.
The conditions for the coexistence of the supersolid with 
the homogeneous superfluid are discussed, 
and the surface tension 
of a representative solid-liquid interface is calculated.
\end{abstract} 
\date{\today}
\pacs{}
\maketitle

\section{\bf INTRODUCTION} 

A supersolid is a phase of matter
that shows both crystalline and superfluid properties,
i.e. the simultaneous breaking of continuous translational
and global gauge symmetries, as originally proposed in Ref.\cite{gross},
resulting in the formation of an ordered crystal structure
with a phase coherence that allows for partial superfluid flow
through the solid \cite{chester,leggett}.

Theoretical and experimental efforts have 
been focused in recent years to investigate
the most natural candidate for supersolidity, 
i.e. $^4$He at low temperature, especially after the
apparent observation of Non-Classical 
Rotational Inertia (NCRI) effects by Kim and Chan \cite{chan}.
Consensus is lacking, however, as to whether 
the experimental data
are really a manifestation of 
supersolidity in $^4$He, and more
recent measurements cast some doubts on this
hypothesis \cite{chan1} (for a recent review about possible
supersolidity in $^4$He and other system, see Ref.\cite{rev,boninsegni}). 

The possibility of formation of a solid structure
simultaneously possessing crystalline and superfluid
properties 
has been associated long ago \cite{russi} with an
excitation spectrum of the liquid phase characterized by a roton
minimum at finite $q$-vector, the liquid to "supersolid" transition 
being triggered by softening of the roton minimum.

A necessary condition for developing a roton minimum
is that the interaction
pair-potential has a Fourier transform that 
becomes negative in some range of the q-vector. 
This can be understood by recalling 
the Bogoliubov dispersion for the uniform liquid
of bosonic particles interacting via a pair potential $V(r)$
\beq
\hbar \omega (q)= \sqrt{({\hbar^2q^2\over 2M})
[{\hbar^2q^2\over 2M} +2\rho \tilde {V}(q)] }
\label{disp}
\eeq
where $\tilde {V}(q)$ is the Fourier transform 
of the interaction potential $V(r)$ and $\rho $ is the liquid density.
When $\tilde {V}(q)$ has a 
negative contribution in some range of q-vectors,
thus balancing the quadratic q-term in the above expression,
a roton minimum may develop, and
the roton gap decreases with increasing the density,
ultimately vanishing at some critical value 
where $\omega $ becomes imaginary. 
This marks the onset of 
a dynamical instability at which density 
modulations may spontaneously develop with no
energy cost. 
Such softening instability could be equivalently reached 
in a flowing superfluid with a non-zero roton gap, 
as predicted for $^4$He flow \cite{pita}, 
and later seen in Density Functional simulations \cite{anci}.
A similar effect arises
in metastable superflow states of soft-core bosons in 2D \cite{kumi}.

If a roton minimum is present, spontaneous 
solidification into a crystal structure is actually possible
even before the roton gap disappears \cite{russi,pomeau}.
As numerically found, for instance,  
in Ref.\cite{pomeau}, by increasing the density, 
the roton gap
decreases until a critical value is reached where the
system undergoes a first-order phase 
transition to an ordered structure,
which {\it may} have supersolid nature. 
This happens for instance in the case of $^4$He where, by applying
pressure, the roton gap decreases because of the associated 
increase in the density, and the formation of a crystal 
occurs well before the roton gap disappears \cite{rossi1}. 

Supersolid phases have been recently predicted 
for confined condensed spinless bosons in 2-dimensions 
interacting via a broad class 
of soft core repulsive potentials
(i.e. short range interactions which
do not grow arbitrarily large at short distance 
but remain instead at a finite value, say $V_0$), 
which might be experimentally realized
via strongly enhanced Van der Waals 
interaction between Rydberg atoms \cite{henkel,cinti},
making them ideal ingredients 
to realize exotic quantum many-body phases of 
matter \cite{cinti,schauss,pupillo,henkel,rydberg}.
At low temperature, such systems are predicted to 
make a transition from the condensed superfluid phase 
to a crystal structure made of atomic
clusters arranged in an ordered lattice 
superstructure, resulting in a number 
of supersolid phases \cite{kivelson,cinti}, the
phase coherence being established through quantum
hopping of atoms across adjacent clusters.

Classically, cluster formation comes as a 
consequence of the soft-core interaction, 
i.e. the energy cost for forming close particle pairs is
bound by $V_0$, so that the energy cost associated with 
the particle overlap remains finite.
This may enable the formation of "solid" structures made of clusters
of atoms
when the density is so high that multiply occupied states 
become energetically favored upon increasing the lattice
constant.
In the associated quantum bosonic system, two important 
effect are added, i.e.
the possibility of particle "hopping" between 
adjacent clusters and the possibility of exchange-driven 
Bose-Einstein condensation, resulting in supersolidity.

The zero-temperature phase diagram of 
two-dimensional Bosons with finite range
soft-core interactions has been theoretically 
described 
in Ref.\cite{cinti1,saccani}, where 
both mean-field and first principles
Monte Carlo simulations have been used.
This model is known to admit in 2-dimensions 
a ground state which is a "cluster crystal",
and moreover it supports NCRI;
three sound (gapless) modes are expected in the solid phase, 
identifiable as the Goldstone bosons associated with 
the breaking of continuous global symmetries.
Besides the two phonon branches
of a "normal" two-dimensional crystal associated with the 
breaking of translational invariance in 2-dimensions, 
a third mode may appear, associated
with the breaking of global gauge invariance which 
leads to the phase coherence of a superfluid fraction.
Quantized vortices are also expected in a supersolid, 
just like in the superfluid phase,
as shown in the numerical simulations of Ref.\cite{rydberg}.
A supersolid can thus be defined as any inhomogeneous structure with
translational long-range order, which exhibits an 
excitation spectrum with the characteristic 
features described above \cite{rydberg}.

The excitation spectrum of a soft-core bosonic 
supersolid has also been studied 
using first principles method in Ref.\cite{saccani1}
for a 2-dimensional supersolid structure,
showing both the phonon modes appropriate to a solid structure
and a softer collective excitation related to broken
translational and gauge symmetry, respectively.

In Ref.\cite{macri} the excitation spectrum of a 2-dimensional 
soft-core supersolid has been computed at a mean-field level 
by solving the 
Bogoliubov-deGennes equations, clearly showing the presence of the 
mode associated with the superfluid fraction
of the solid. In particular, a rather good agreement was found
between the mean-field predictions
and the results of Quantum Monte Carlo simulations \cite{macri,saccani1}.

In Ref.\cite{li} a Bose-Einstein supersolid phase in 2-dimensions
with a step-like pair interaction showing a 
soft-core below a given core radius and
a dipolar repulsive term at larger distances has been studied
and the phase diagram obtained.
The nucleation of vortices in a dipole-blockade 2-dimensional 
supersolid condensate and its effect on the superfluid fraction
has been investigated in Ref.\cite{mason}.
Again, evidence of a supersolid phase of a 2-dimensional dipolar 
crystal has been found from zero-temperature Quantum
Monte Carlo simulations \cite{boro}. 

All of the theoretical results 
mentioned above are in 2-dimensions except for the 
mean-field study of the ground-state
structural properties of the 3D nonuniform phase of
bosons interacting through a dipole-blockade type 
interaction $V(r)=C_6/(r^6+R_c^6)$ \cite{henkel}, 
which constitutes a strong candidate to exhibit supersolid 
properties. In a recent work Saccani et al \cite{saccani1} 
show that, in 2D, the simple model of "soft spheres" 
features the same basic physics as produced by interaction 
with a $1/r^n$ interaction outside the soft sphere. 

It seems thus timely to investigate in a 
quantitative way, even if only
at the mean-field level, a three-dimensional system of 
bosons interacting through a soft-core potential, 
to establish the range
of values of the interaction strength
that favors a crystal structure made of atomic nano-clusters,
and to study both the static properties
of the supersolid structure and its excitation spectrum.
Unlike the case of 2D soft-core boson supersolids 
where both mean field \cite{rydberg,macri}
and Monte Carlo calculations \cite{macri,saccani1} have been performed, 
no ab-initio results exist for the three-dimensional system
which may serve as a benchmark to assess the accuracy of
the present calculations. For this reason we have
performed some preliminary path integral ground 
state (PIGS) Monte Carlo simulations of the 3D system to substantiate 
our mean field findings.
On the basis of the excellent agreement between mean-field and ab initio
methods for the 2D system of soft-core bosons \cite{macri}, both for the 
critical point of the superfluid-supersolid transition and for the 
dispersion relations of collective excitations, and on the 
preliminary results of our 
PIGS simulations, we expect that our calculations are accurate enough to
provide a reliable picture which might be helpful to understand the 
properties of this prototype system in preparation of experiments in the
near future. 

In the following,
we try to understand some basic questions such as: 
a) which is the ground state structure of the supersolid? 
b) how large a number of particles per unit cell is requested to 
realize the supersolid, given the interaction parameters? 
c) how does the superfluid fraction in the supersolid 
phase depend on the interaction parameter? 
d) is it possible to have coexistence of the supersolid and the superfluid? 
and what is the solid-liquid interface energy?

\section{\bf METHODS AND CALCULATIONS} 

We consider $N$ bosonic atoms of mass $M$ 
interacting through a pair-potential
$V(r)$ represented by a simple "soft-sphere" interaction:
\beq
V(r)=V_0 \Theta (R_c-r)
\label{ss_pot}
\eeq
where $V_0$ and $R_c$ are the height and width of the 
potential ($\Theta $ is the Heavyside step function 
and $r$ is the length of the 3D vector $\bf{r} \equiv (x,y,z))$.
While the above interaction does not occur naturally 
the so-called Rydberg dressing \cite{henkel,pupillo}
of atomic BEC constitutes, as 
described in the previous Section, 
a promising approach for an experimental realization of 
this interaction.
The actual potential in Rydberg-blockade interaction 
has an additional long-range Van der Waals repulsive tail which decays
rapidly with distance, 
rather than being abruptly cut as in the simpler soft-core model.
However,  on the basis of numerical
simulations \cite{saccani,macri} it has been shown
that the formation of supersolid is 
largely insensitive to the actual shape of the 
repulsive interaction, provided it produces a roton
minimum in the dispersion relation.

We assume that all the $N$ atoms of the system 
are in a Bose-Einstein condensate 
described by the wavefunction $\Phi ({\bf r})$. 
The energy of the system at the mean-field level 
is thus expressed by the functional
\begin{eqnarray}
E[\Phi ,\nabla \Phi]={\hbar ^2 \over 2M}\int {|\nabla 
\Phi ({\bf r})|^2}d{\bf r}+\nonumber \\
{1\over 2}\int {\int {V(|{\bf r}-
{\bf r}^\prime |)|\Phi ({\bf r})|^2  }
|\Phi ({\bf r}^\prime )|^2d{\bf r}\,d{\bf r}^\prime}
\end{eqnarray}
\label{energy}
Functional minimization of the above energy
leads to the following Euler-Lagrange equation
\beq
\hat {H}\Phi \equiv {\Big [} -\frac{\hbar^2 }{ 2M}\nabla ^2  +
\int{V(|{\bf r}-{\bf r}^\prime|)}
|\Phi({\bf r}^\prime )|^2d{\bf r}^\prime {\Big ]}\Phi({\bf r}) = \mu \Phi({\bf r})
\label{schrod}
\eeq
where $\hat {H} $ is defined by the terms in square brackets,
and
where $\mu $ is a Lagrange multiplier whose value is determined
by the normalization condition $\int |\Phi ({\bf r})|^2 d{\bf r}=N$.

This equation will be solved numerically,
as explained in the next Section, to yield the lowest 
energy state $\Phi $ describing the condensate
in the ground-state.

Upon scaling lengths by $R_c$ and 
energies by $\hbar ^2/MR_c^2$, and
once the wave function is scaled as 
$\Phi /\sqrt{\rho }$, the above model has 
a single dimensionless parameter $\Lambda $
that determine the solutions of Eq.(\ref{schrod}). 
In 3D it reads:
\beq
\Lambda \equiv M\rho V_0R_c^5/\hbar ^2 .
\label{lambda}
\eeq
$\Lambda $ can be varied, e.g., by changing the density of the system,
although from the experimental 
point of view the optimal control parameter 
is $V_0$, which can be easily varied since it depends strongly
on the quantum number of Rydberg states \cite{henkel,pupillo}.

The mean-field equation (\ref{schrod}) is 
valid when the quantum fluctuations in the region
inside the range of the potential are relatively small, which occurs
when the average particle number inside this range is large.
In the present case this means that
the number of particles within each cluster 
forming the solid structure should be large.
However, as shown in Ref.\cite{macri}, 
the mean-field description turns out to be rather accurate 
even when this number is relatively small
(of the order of a few atoms).

In order to compute the excitation spectrum, we make 
the usual Bogoliubov transformation to a 
Hamiltonian describing a collection of non-interacting
quasi-particles for which the condensate is the vacuum:
\beq
\Psi ({\bf r},t)=e^{-i\mu t/\hbar}
[\Phi ({\bf r})+u_{n,{\bf k}}({\bf r})e^{-i\omega t}
-v^\ast _{n,{\bf k}} ({\bf r})e^{i\omega t} ]
\label{bogo}
\eeq
where $u_{n,{\bf k}}({\bf r})$ and $v_{n,{\bf k}}({\bf r})$ 
are the wavefunctions of the excitation mode with band 
index $n$ and wavevector ${\bf k}$ and $\Phi ({\bf r})$ 
is the solution of Eq. (\ref{schrod}).
Substituting this form into the
time-dependent Schrodinger equation:
\beq
i \hbar \frac{\partial}{\partial t}\Psi({\bf r}) = 
{\Big [} -\frac{\hbar^2 }{ 2M}\nabla ^2  +
\int{V(|{\bf r}-{\bf r}^\prime|)}
|\Psi({\bf r}^\prime )|^2d{\bf r}^\prime {\Big ]}\Psi({\bf r})
\label{tdschrod}
\eeq 
associated with the Hamiltonian in Eq.(\ref{schrod})
and keeping only terms
linear in the functions $u,v$, one obtains 
the following coupled equations:
\begin{eqnarray}
\hbar \omega u_{n,{\bf k}}({\bf r})=
{ \big [}-\frac{\hbar^2 }{ 2M}\nabla ^2-\mu+\int { V(|{\bf r}-{\bf r}^\prime|)
|\Phi ({\bf r}^\prime )|^2 d{\bf r}^\prime }{\big ]}
u_{n,{\bf k}}({\bf r}) \nonumber \\
+\Phi({\bf r})\int { V(|{\bf r}-{\bf r}^\prime|)
\Phi ^\ast ({\bf r}^\prime)
u_{n,{\bf k}}({\bf r}^\prime)d{\bf r}^\prime }\nonumber \\
-\Phi({\bf r})\int { V(|{\bf r}-{\bf r}^\prime|)
\Phi ({\bf r}^\prime)
v_{n,{\bf k}}({\bf r}^\prime)d{\bf r}^\prime } \,\,\,
\nonumber \\
-\hbar \omega v_{n,{\bf k}}({\bf r})={\big [}-\frac{\hbar^2 }{ 2M}\nabla ^2 -\mu 
+\int { V(|{\bf r}-{\bf r}^\prime|)
|\Phi ({\bf r}^\prime )|^2d{\bf r}^\prime {\big ]}
v_{n,{\bf k}} }({\bf r}) \nonumber\\
-\Phi ^\ast({\bf r})\int { V(|{\bf r}-{\bf r}^\prime|)
\Phi ^\ast ({\bf r}^\prime)
u_{n,{\bf k}}({\bf r}^\prime)d{\bf r}^\prime }\nonumber \\
+\Phi ^\ast ({\bf r}) \int { V(|{\bf r}-
{\bf r}^\prime|)\Phi ({\bf r}^\prime)
v_{n,{\bf k}}({\bf r}^\prime)d{\bf r}^\prime } \,\,\,
\label{bdgeq}
\end{eqnarray}
We expand the (real) function $\Phi $ and
the complex functions $u,v$ in the Bloch form 
appropriate to a periodic system:
\begin{eqnarray}
\Phi ({\bf r})=\sum _{\bf G} \Phi_{\bf G} 
e^{i\bf G \cdot \bf r} \\
u_{n,{\bf k}}({\bf r})=e^{i\bf k \cdot \bf r}\sum _{\bf G} 
u_{\bf k+\bf G}^{(n)} e^{i\bf G \cdot \bf r} \\
v_{n,{\bf k}}({\bf r})=e^{i\bf k \cdot \bf r}\sum _{\bf G} 
v_{\bf k+\bf G}^{(n)} e^{i\bf G \cdot \bf r}
\label{bloch}
\end{eqnarray}
In the above expansions, the ${\bf G}$-vectors are the
reciprocal lattice vector appropriate to the 
space symmetry of the cluster-crystal structure.
By making the above substitutions
in Eq.(\ref{bdgeq}) (and omitting the band index $n$ for clarity) one gets:
\begin{eqnarray}
[{\hbar ^2\over 2M}({\bf k}+{\bf G})^2-\mu -\hbar \omega]u_{\bf k +\bf G}\nonumber \\
+\sum _{\bf G^ \prime} \tilde {U}_{\bf G-\bf G^ \prime}u_{\bf k +\bf G^ \prime}
+\sum _{\bf G^ \prime,\bf G^ {\prime \prime}}\Phi _{\bf G^ {\prime \prime}-\bf G^ \prime}
\Phi _{\bf G-\bf G^ {\prime \prime}}\tilde {V}_{\bf k+\bf G^ {\prime \prime}} 
u_{\bf k+\bf G^ \prime }\nonumber \\
-\sum _{\bf G^ \prime,\bf G^ {\prime \prime}}\Phi _{\bf G^ {\prime \prime}-\bf G^ \prime}
\Phi _{\bf G-\bf G^ {\prime \prime}}\tilde {V}_{\bf k+\bf G^ {\prime \prime}} 
v_{\bf k+\bf G^ \prime }=0 \nonumber \\
\medskip
-[{\hbar ^2\over 2M}({\bf k}+{\bf G})^2-\mu +\hbar \omega]v_{\bf k +\bf G}\nonumber \\
-\sum _{\bf G^ \prime} \tilde {U}_{\bf G-\bf G^ \prime}v_{\bf k +\bf G^ \prime}
-\sum _{\bf G^ \prime,\bf G^ {\prime \prime}}\Phi _{\bf G^ {\prime \prime}-\bf G^ \prime}
\Phi _{\bf G-\bf G^ {\prime \prime}}\tilde{V}_{\bf k+\bf G^ {\prime \prime}} 
v_{\bf k+\bf G^ \prime }\nonumber \\
+\sum _{\bf G^ \prime,\bf G^ {\prime \prime}}\Phi _{\bf G^ {\prime \prime}-\bf G^ \prime}
\Phi _{\bf G-\bf G^ {\prime \prime}}\tilde{V}_{\bf k+\bf G^ {\prime \prime}} 
u_{\bf k+\bf G^ \prime }=0 \nonumber \\
\label{bdgeq_g}
\end{eqnarray}
$\tilde {V}_{\bf q} $ is the 
Fourier transform of the soft-core interaction:
\beq
\tilde {V}_{\bf q} \equiv \int {V(r) e^{i {\bf q}\cdot {\bf r} }}
d{\bf r}=4\pi V_0 R_c^2 j_1(qR_c)/q \,
\label{v_g}
\eeq
where $j_1(x)=sin(x)/x^2-cos(x)/x$ is 
the spherical Bessel function of the
first kind.

The quantities $\tilde {U}_{\bf G}$ in Eqns.(\ref{bdgeq_g}) 
are defined through 
\beq
\int{ V(|{\bf r}-{\bf r} ^\prime|) |
\Phi ({\bf r} ^\prime)|^2d{\bf r} ^\prime }=
\sum _{\bf G} \tilde {U}_{\bf G}
e^{i{\bf G}\cdot {\bf r}}
\label{utilde}
\eeq

It is useful to introduce the following matrices
(with dimensions $(n_r^3\times n_r^3)$,
where $n_r$ is the real space mesh
used to integrate the stationary equation, 
see the following Section): 
\begin{eqnarray}
{\bf A}_{{\bf G},{\bf G}^\prime}
\equiv \delta _{{\bf G},{\bf G}^\prime}[{\hbar ^2
\over 2M}({\bf k}+{\bf G})^2-\mu 
]\nonumber \\
+\tilde {U}_{\bf G-\bf G^ \prime}
+\sum _{\bf G^ {\prime \prime}}\Phi _{\bf G^ 
{\prime \prime}-\bf G^ \prime}
\Phi _{\bf G-\bf G^ {\prime \prime}}\tilde{V}_{\bf k+
\bf G^ {\prime \prime}}
\end{eqnarray}
\begin{eqnarray}
{\bf B}_{{\bf G},{\bf G}^\prime}
\equiv -\sum _{\bf G^ {\prime \prime}}
\Phi _{\bf G^ {\prime \prime}-\bf G^ \prime}
\Phi _{\bf G-\bf G^ {\prime \prime}}
\tilde{V}_{\bf k+\bf G^ {\prime \prime}} 
\end{eqnarray}

The system (\ref{bdgeq_g}) can thus be written in the matrix form:
\begin{equation}\everymath{\displaystyle}
\begin{bmatrix}
{\bf A} & {\bf B}  \\
-{\bf B}  & -{\bf A}  \\
\end{bmatrix}\begin{pmatrix} {\bf u} \\ {\bf v}
\end{pmatrix}=\hbar \omega
\begin{pmatrix} {\bf u}
\\ {\bf v}   \,
\end{pmatrix}
\end{equation}

The excitation frequencies $\omega ({\bf k})$ 
can be determined from the solutions 
of the above non-Hermitian eigenvalue problem.
This can be reduced to a non-Hermitian problem 
of {\it half} the dimension 
(thus largely reducing the computational cost
of diagonalization)
by means of a unitary transformation \cite{unitary}:
\beq
({\bf A}-{\bf B}) ({\bf A}+{\bf B}) |{\bf u}
+{\bf v}>=(\hbar \omega )^2
|{\bf u}+{\bf v}>
\label{unitary}
\eeq

If needed, one may calculate the separate ${\bf u}$, ${\bf v}$ 
by properly combining the
eigenvectors of Eq. (\ref{unitary}) with those 
of the associated eigenvalue problem
\beq
({\bf A}+{\bf B}) ({\bf A}-{\bf B}) |{\bf u}-{\bf v}>=(\hbar \omega )^2
|{\bf u}-{\bf v}> ,
\label{unitary1}
\eeq
again of reduced dimensions.

In order to provide a quantitative benchmark for the mean field results
presented here,
we have performed also some preliminary 
path integral ground state (PIGS) Monte Carlo
simulations \cite{pigs}.
PIGS is an {\it exact} $T = 0$ K method that allows to obtain the ground
state of a given microscopic Hamiltonian by projecting in imaginary time 
a trial wave function.
The PIGS method is unbiased by the choice of such a trial wave function 
and the only inputs are the interparticle potential and the approximation
for the imaginary time propagator \cite{rossi2}.
Here we consider a system of $N$ Bosons in a cubic box with periodic
boundary conditions interacting via the soft-core potential 
(\ref{ss_pot}); we project a constant wave function and 
we consider the Suzuki pair approximation for the imaginary time 
propagator \cite{rossi2}.
Despite the possibility of reducing the computational cost of PIGS
simulation by choosing a trial wave function as close as possible to the
ground state of the system, Quantum Monte Carlo simulations 
of the 3-dimensional system studied here are still much 
more computationally expensive than mean field methods.
Encouraged by the consistency of our preliminary PIGS results with the mean 
field ones, we choose to perform only a limited number of calculations 
with the aim to give support to the present mean field study, leaving a 
systematic full Quantum Monte Carlo investigation of 3D soft Boson 
systems for a future work.  

\section{\bf RESULTS AND DISCUSSION} 

\begin{figure}
\epsfig{file=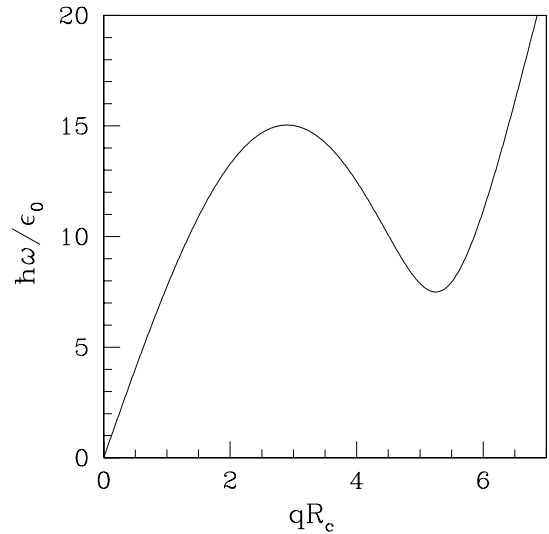,height=3.8 in,clip=}  
\caption{Dispersion relation for the homogeneous system, for $\Lambda =15.8$.
Energies are expressed in units of 
$\epsilon _0\equiv \hbar ^2/(MR_c^2)$. }
\label{fig1}
\end{figure}

In Fig.\ref{fig1} the dispersion relation (\ref{disp}) for the 
uniform liquid is shown for a value of the 
interaction strength $\Lambda =15.8$.
Energies are expressed in units of $\epsilon _0=\hbar ^2/(MR_c^2)$.
For
sufficiently large values of $\Lambda$, $\tilde{V}(q)$ has a 
negative contribution around $q_{rot}\sim 2\pi /R_c$ and a roton minimum appears.
The roton gap decreases upon increasing $\Lambda$ 
until a threshold value
$\Lambda _r$ is reached 
where Landau's critical velocity (defined as the tangent at the 
roton minimum) becomes zero, i.e. when 
\beq
(\rho M/q)dV(q)/dq=-1/2
\label{inst}
\eeq
This condition (closing of the roton gap) marks the onset of 
a roton instability at which density modulations may develop without
energy cost, as discussed in the Introduction.
The condition (\ref{inst}) can be recast in a form involving only 
the adimensional strength parameter $\Lambda $, giving
\beq
\Lambda _r =21.71
\label{lambdac}
\eeq
This value could be achieved, e.g., with the 
following choice of parameters: $^{87}$Rb
condensate of density $2.2 \times 10^{-11}\,a_0^{-3}$, with
$R_c=6000 \,a_0$ and $V_0= 183\,nK$.

As shown in the following, and similarly to what occurs
in the 2-dimensional case \cite{cinti,macri,saccani},
the homogeneous liquid phase destabilizes 
well before the roton instability condition (\ref{lambdac}) is reached,
spontaneously converting the liquid phase 
into an ordered solid-like structure
("cluster crystal").

\subsection{\bf Ground-state properties} 

We have numerically solved Eq.(\ref{schrod}) by
propagating it in imaginary time, i.e. by solving the equation
\beq
{\partial \Phi \over \partial t}+(\hat {H}-\mu )\Phi ({\bf r}) = 0
\label{imag}
\eeq
The wave function $\Phi({\bf r})$
is represented on a three-dimensional uniform mesh in real space,
with periodic boundary conditions imposed on the system.
By studying the convergence in energy of the 
solution with increasing number of
points in the mesh, we verified that a relatively 
coarse grid made of $n_r^3$ points with $n_r=20$
is sufficient to accurately describe $\Phi({\bf r})$.
To compute the spatial derivatives
appearing in the previous equation, 
we used a 11-point finite-difference formula \cite{pi}.
The convolution integral in the potential energy term 
of Eq.(\ref{schrod}) is 
efficiently evaluated in  reciprocal space 
by using Fast Fourier transform techniques.

Depending on the value of $\Lambda $, the lowest
energy structure described by $\Phi({\bf r})$
is either a uniform liquid or
a structured system with long-range order.
For sufficiently large values of $\Lambda $, in fact,
the system spontaneously breaks the translational invariance leading
to the appearance of a crystalline phase made of individual
clusters of atoms arranged in an ordered structure.
This phase, as discussed in the following, 
can exhibit a supersolid behavior by decoupling
a superfluid component from the crystalline structure.

We have studied the relative stability of the liquid vs.
solid phase as a function of the interaction parameter $\Lambda $.
We find that, upon increasing $\Lambda $, 
once a critical value 
\beq
\Lambda _{ss} = 15.2
\label{ss}
\eeq
is reached and well before the roton instability 
(i.e. the disappearance of the roton gap) occurs,
the structure spontaneously converts into an ordered arrangement
of clusters, each containing a number of atoms that increases as
$\Lambda $ is further increased. 

\begin{figure}
\epsfig{file=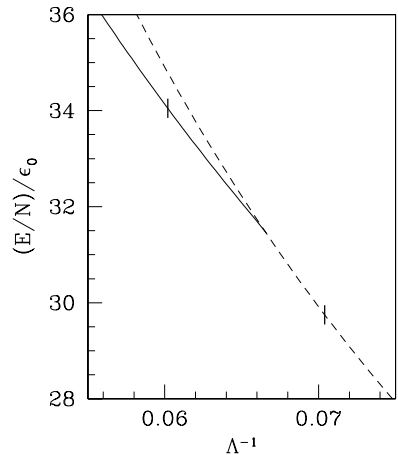,height=3.3 in,clip=}  
\caption{Energy/atom for the liquid and solid phases, shown as
a function of $\Lambda ^{-1}$. Energies are in 
units of $\epsilon _0\equiv \hbar ^2/(MR_c^2)$.
Dashed line: liquid phase; solid line: FCC cluster crystal.
The two vertical ticks show the two values of $\Lambda $
at coexistence as obtained by a double-tangent construction 
(see the text).
}
\label{fig2}
\end{figure}

To assess the relative stability of different possible solid phases
we have studied different crystal structures (SC, FCC and HCP) 
and optimized, for each 
structure, the associated lattice parameter for a given value 
of $\Lambda $. We find that face-centered-cubic (FCC) and 
hexagonal-close-packed (HCP) ordering are almost degenerate
in energy, with the FCC structure only slightly favored.
We have checked, by using a finer mesh than
the one we eventually used to compute the
excitation spectrum, that the FCC structure is indeed 
always (slightly) favored over the HCP one 
in a wide range of values of the 
interaction parameter $\Lambda $.
The FCC  arrangement is the one produced also by PIGS computations.
This is not surprising, due to the 
short range nature of the interactions involved, 
since FCC and HCP structures differ only in 
the second nearest-neighbor atoms arrangement.
This is similar to what occurs 
in the three-dimensional boson system interacting through
the Rydberg blockade type interaction \cite{henkel} 
which contains also a long range repulsion, where
again FCC ordering has the lowest possible energy per atom. 

We compare the energy per atom (expressed in units of
$\epsilon _0\equiv \hbar ^2/(MR_c^2)$) of the liquid and solid
phases in Fig.\ref{fig2}. 
The liquid-solid transition 
occurs at $\Lambda _{ss}=15.2$. For higher values, the solid structure 
is always favored. We plot $\epsilon $ as a function of $\Lambda ^{-1}$
rather than of $\Lambda $
for a reason that will be clarified in the next Subsection.
Our preliminary PIGS simulations locate the 
liquid--solid transition between $\Lambda = 15.0$ and $\Lambda = 15.5$,
in good agreement with our mean-field results.

\begin{figure}
\epsfig{file=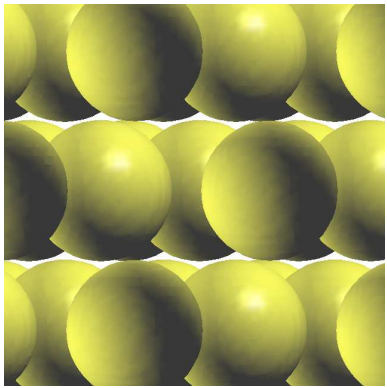,height=2.0 in,clip=}  
\caption{(Color online) Equilibrium equidensity plot for the 
$\Lambda =15.8$ FCC cluster solid.
Each cluster contains approximately 10 atoms.
The surface of equal density is drawn at the value 
$ \rho  /2$.}
\label{fig3a}
\end{figure}

Note that the energy per atom is an increasing function of 
$\Lambda $
(i.e. of the density, for a given $V_0$ and $R_c$), which is
an obvious consequence of the repulsive character of the 
interparticle interaction: therefore all the phases discussed here, with a finite density, 
are meaningful only in confined systems, a condition which is 
certainly realized in cold gases experiments.

\begin{figure}
\epsfig{file=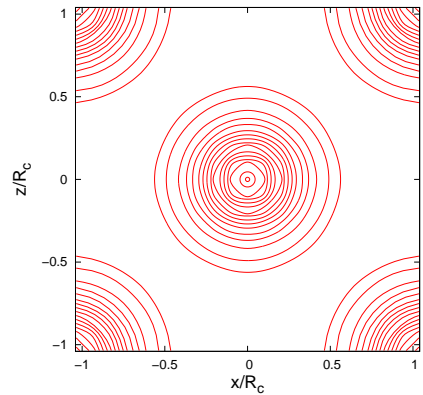,height=3.3 in,clip=}  
\caption{Equilibrium density contour map for the 
$\Lambda =15.8$ FCC cluster solid
in the (100) plane, shown in the conventional cubic cell containing
four clusters. Each cluster contains approximately 10 atoms.
The inner contour line is drawn at the value $16 \rho $ ($\rho =N/\Omega $
being the density of the liquid phase), while the external contour line
are drawn at the value $ \rho  /2$.}
\label{fig3}
\end{figure}

As an example, in Fig.\ref{fig3a} we show the 
resulting 3-dimensional density plot for 
the optimized FCC cluster solid obtained
with $\Lambda =15.8$.
For clarity, we also show in Fig.\ref{fig3}
the density profile by means of 
contour lines in the $(100)$ plane. The cell shown is the conventional
cubic cell for FCC structures, containing four equivalent sites.
Each cluster contains about 10 atoms.
The equilibrium interatomic distance is $d\sim \rho ^{-1/3}=0.595\,R_c$,
while the cluster-cluster distance is $d_c=1.465\,R_c $.
For comparison, the characteristic wavelength 
associated with the roton minimum in the dispersion 
relation for the homogeneous system is $2\pi/q_{min}=1.208\,R_c$.

By increasing $\Lambda $ (i.e. by increasing 
the atomic density at constant $V_0$ and $R_c$), 
the number of atoms inside 
each cluster grows, while the overlap between 
adjacent cluster decreases, 
suggesting that the superfluid fraction (which is associated
to a global coherence maintained by hopping of atoms between 
adjacent clusters) is also decreasing with increasing the density.
This will be confirmed by the calculation of the superfluid
fraction, as shown in the following.

\begin{figure}
\epsfig{file=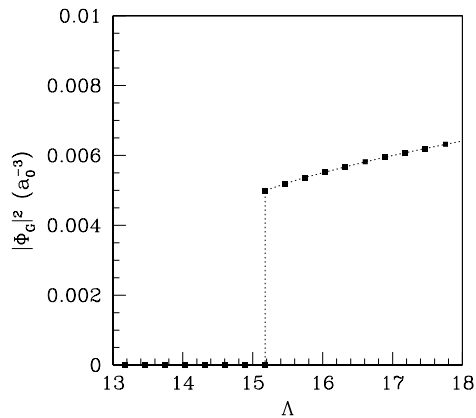,height=3.0 in,clip=}  
\caption{Occupation fraction of the 
lowest finite momentum state (see the text).}
\label{fig4}
\end{figure}

The liquid-solid transition 
which occurs at $\Lambda _{ss}=15.2$ is first-order
in character, being 
accompanied by a discontinuity in the derivative of the
energy per atom. The nature of the transition is 
better appreciated by looking at the
occupation fraction of the lowest finite momentum, i.e.
the lowest G-vector component $\Phi _{\bf G}$ 
of the wavefunction $\Phi ({\bf r})$
corresponding to 
the FCC reciprocal lattice vector $4\pi /a_{ss}$
($a_{ss}=2.07\,R_c$ being the equilibrium lattice 
constant of 
the cubic structure shown in Fig.(\ref{fig3})).
Fig.\ref{fig4} clearly shows the 
first-order jump in $\Phi _{\bf G}$ as the
transition value $\Lambda  _{ss}$ is reached.

From the dependence of the energy 
per atom $\epsilon\equiv E/N$
on the atomic density one can compute, e.g.,  
the inverse of thermodynamic compressibility, as 
\beq
\kappa ^{-1}= \rho \Lambda {\partial \mu \over \partial \Lambda}
\label{compress}
\eeq
where the chemical potential $\mu $ is computed in turn
from the data shown in Fig.(\ref{fig2}) as:
\beq
\mu = {\partial E \over \partial N} =\epsilon + 
\rho {\partial \epsilon \over \partial \rho}=\epsilon (\Lambda )+
\Lambda {\partial \epsilon \over \partial \Lambda}
\label{chem_pot}
\eeq

\begin{figure}
\epsfig{file=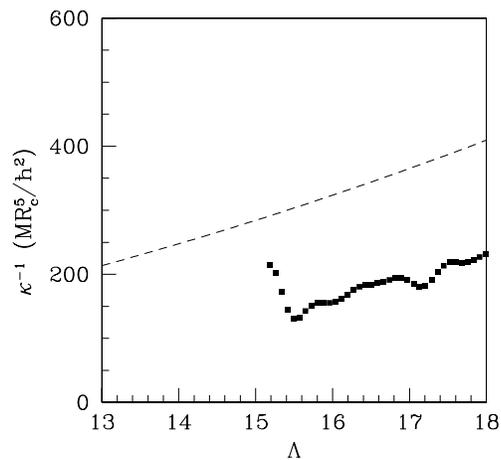,height=3.3 in,clip=}  
\caption{Inverse compressibility for the case 
$\rho =2.2\times 10^{-11} a_0 ^{-3}$: homogeneous liquid (dashed line);
cluster crystal (filled squares).}
\label{fig5}
\end{figure}

We show the resulting inverse compressibility 
in Fig.(\ref{fig5}), for 
$\rho =2.2\times 10^{-11} a_0 ^{-3}$.
A drop occurs at the liquid-solid transition,
showing that the cluster solid is remarkably softer than 
the liquid phase at the same density.
The inverse compressibility is related to the low-q 
average sound velocity $c$ through the relation
\beq
\kappa ^{-1}= \rho M^\ast c^2
\label{sound}
\eeq
where $M^\ast $ is some effective mass due to the periodic potential.
The drop in Fig.(\ref{fig5}) is thus 
associated to an overall softening of the
sound velocity. As we will show in the following, this
behavior is associated with the presence, in addition to the
usual phonon modes, of a soft Bogoliubov mode which 
is the signature of the supersolid phase.

\subsection{\bf Superfluid fraction} 

Because of its condensate character, the cluster crystal
described above
can naturally exhibit superfluidity, resulting in a nonzero
Non-Classical Rotational Inertia (NCRI) \cite{leggett}.
Numerical proof that this is indeed the case for soft-core
bosons has been provided in Ref.\cite{afta}.
Equivalently \cite{sepulveda}, one can instead consider 
the momentum of the crystal under a Galilean boost ${\bf v}$
and solve the associated time-dependent Schrodinger equation in the
co-moving frame of reference:
\beq
i\hbar {\partial \over \partial t} \Psi = \Big[\hat{H}+i\hbar {\bf v}
\cdot \nabla \Big]\Psi
\label{nlse1}
\eeq
By computing the linear momentum of the system one can define
the superfluid fraction $f^{ss}$ as the tensor \cite{sepulveda1}
\beq
f^{ss}_{ik}=\delta _{ik}-\lim _{|v|\rightarrow 0} 
{1  \over N}
{\partial P_i \over \partial v_k }
\label{fraction}
\eeq
where ${\bf P}=-(i\hbar /2)\int {(\Psi ^\ast \nabla 
\Psi -\Psi \nabla \Psi ^\ast)d{\bf r}}$
is the total momentum.

\begin{figure}
\epsfig{file=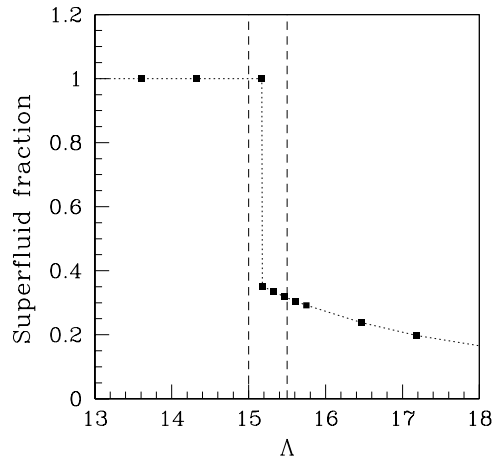,height=3.3 in,clip=}  
\caption{Calculated superfluid fraction as a function of $\Lambda $.
The vertical dashed lines indicate the $\Lambda$ range for the liquid-solid
transition as obtained by {\it exact} PIGS Monte Carlo method.
}
\label{fig_ss}
\end{figure}

We show in Fig.(\ref{fig_ss}) our calculated values 
for the diagonal part of this tensor along the 
direction of the velocity boost ${\bf v}$.
A sudden, finite drop from $1$ to $\sim 0.4$ occurs at the
liquid-supersolid transition, while the superfluid fraction 
decreases in a monotonic way as $\Lambda $ is further increased.
We mention that for the supersolid triangular structure
realized in 2-dimensions by a system of soft-core bosons \cite{macri}, 
the corresponding jump, computed using Path Integral Monte Carlo
methods, is from 1 to $\sim 0.6$.
For comparison, we show with dashed lines the interval of
$\Lambda $ values where our preliminary PIGS Monte Carlo calculations
allow to locate the liquid-solid transition.

Our calculated superfluid fraction are 
probably overestimating to some extent the 
actual superfluid fraction,
since mean-field methods tend to 
overestimate the superfluid fraction compared to the 
predictions of {\it ab initio} Monte Carlo methods \cite{otterlo},
mainly due to the neglect of fluctuations 
that tend to suppress superfluidity.
However, in 3D the effect of quantum fluctuations should be less
than they are, e.g., in 2-dimensional
system of bosons\cite{macri}.

\begin{figure}
\epsfig{file=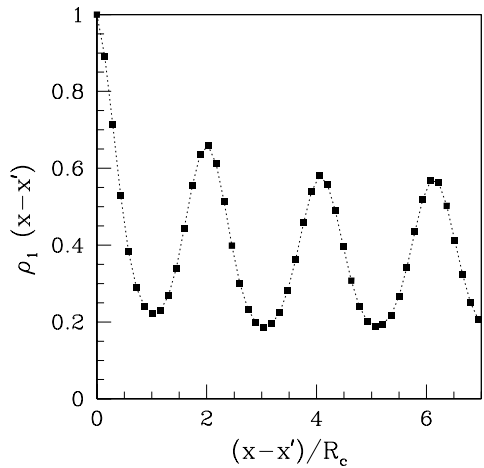,height=3.3 in,clip=}  
\caption{One-body density matrix $\rho_1$ computed with PIGS 
at $\Lambda = 15.72$.
$x$ is along the (100) direction.}
\label{fig_cond}
\end{figure}

The PIGS method does not allow a direct evaluation of the superfluid
fraction, so we cannot directly compare with our mean field results;
however, PIGS provides an estimate of the condensate fraction
$n_0$ \cite{rossi2} whose finite value, at $T=0$ K, is a sufficient
condition for NCRI \cite{leggett2}.
The condensate fraction $n_0$ is obtained as the large distance limit of
the one-body density matrix $\rho_1$.
We have computed $\rho_1$ for $\Lambda = 15.72$, i.e. very close to the
liquid-solid transition, and we find that 
the large distances tail of the one-body
density matrix is converging to a finite value, 
with large superimposed oscillations.
Such oscillations are a typical signature of solid phase, thus
confirming, on one hand, the crystalline structure of the system, but
on the other hand they are so large to prevent us from a precise
evaluation of $n_0$ in systems of tractable sizes.
We show our computed values for $\rho_1$ 
in Fig.(\ref{fig_cond}).

Similarly to what happens in 2D \cite{sepulveda1}
the superfluid fraction decreases exponentially
with the square root of $\Lambda $: $log f^{ss}\propto -\Lambda ^{-1/2}$ 
very close to the transition;
for higher values of $\Lambda $ it
follows instead (approximately) a power law:
$f^{ss}\propto \Lambda ^{-\gamma}$.

\subsection{\bf Supersolid-superfluid coexistence} 

We show here that when 
$\Lambda $ is greater than
$\Lambda _{ss}$, such that a supersolid phase is expected,
but close to $\Lambda _{ss}$,
the spontaneous formation of 
interfaces between solid and liquid patches
might be energetically favored over the realization of a single
extended solid phase.
A similar conclusion was drawn in Ref.\cite{henkel}
based on numerical calculations,
where coexistence between small crystallites and 
liquid embedding them was observed
in a system of bosons interacting through a 
soft-core Rydberg blockade interaction for values 
of the coupling strength very close to the transition point.

The possibility of liquid-solid coexistence in the 
system studied here is proved by a double-tangent construction
on the data shown in Fig.\ref{fig2}. We recall that 
a double-tangent construction on a plot where the
energy per atom $\epsilon $ is drawn as a function of $(1/\rho )$
for the two phases, is equivalent to imposing 
the equality of pressure and chemical potential, 
i.e. the conditions for the thermodynamic 
coexistence of the two phases.
The two vertical ticks in Fig.\ref{fig2} show the 
values of $\Lambda $ characterizing the two coexisting phases
as obtained by such construction, 
$\Lambda _s=16.6$ and $\Lambda _l=14.2$ 
for the solid and liquid phase, respectively.

\begin{figure}
\epsfig{file=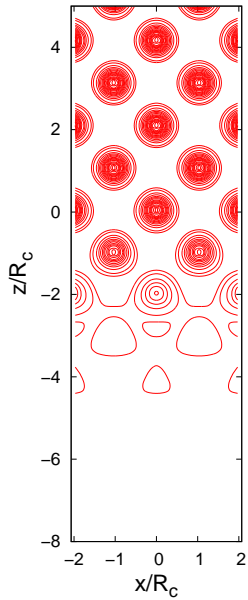,height=3.3 in,clip=}  
\caption{(100) solid-liquid interface for $\Lambda =15.3$.
Only a portion of the supercell used in the calculation 
is displayed for clarity.}
\label{fig6}
\end{figure}

To verify this prediction we have explicitly realized
such two-phase system.
We show in Fig.(\ref{fig6}) the $(100)$ planar interface, 
which we have computed by solving Eq.(\ref{schrod}) 
with $\Lambda =15.3$ (i.e. just above $\Lambda _{ss}$) in a slab
geometry with a fixed total number of atoms $N$ in a
cell of total volume $\Omega $ (such that $\rho =N/\Omega $),
separating a cluster solid phase (upper part
of the Figure) from a coexisting liquid phase.
For the particular interface 
shown in Fig.(\ref{fig6}) the average density in the 
solid phase (far from the interface) $\rho _{s,c}$,
and the density in the 
liquid phase (far from the interface)
$\rho _{l,c} $ correspond to the values of $\Lambda_s $
and $\Lambda_l $
as obtained by the double tangent construction
described above.

One can estimate the 
width of the liquid ($L_l$) and solid ($L_s$) region as
$L_l=(L_s+L_l)(\rho-\rho_{s,c})/(\rho_{l,c}-\rho_{s,c})\,\,$,
$L_s = L-L_l $.

The interface tension can thus be calculated 
(the factor 2 takes into account the presence, in 
a slab geometry, of two solid-liquid interfaces)
\beq
\sigma =[E-(E_s+E_l)]/(2A)
\label{sigma}
\eeq
where $A$ is the area of the transverse 
section of the supercell parallel to the interface,
$E$ is the total energy of the 
configuration shown in Fig.(\ref{fig6}) and
$E_s=\epsilon (\rho_{s,c})L_s$,
$E_l=\epsilon (\rho_{l,c})L_l$
($\epsilon =E/N$ being the energy per atom
shown in Fig.(\ref{fig2})).

In this way we find the  value  
$\sigma_{\{100\}}=300 \,MR_c^4/\hbar ^2$ for the
interfacial energy, i.e. for the energy cost 
to create a $(100)$ solid-liquid interface of unit area.

\subsection{\bf Bogoliubov excitation spectrum} 

The excitation spectrum of the supersolid phase described
above will be discussed next.
As already stated in the Introduction, a consequence of the 
breaking of a global gauge symmetry is the emergence
of a new Goldstone boson, i.e. a gapless mode in the excitation 
spectrum (which we will call Bogoliubov mode in the following), 
in addition to the usual phonon branches.
To check for the appearance of such a mode, we
have computed the frequency 
spectrum of the supersolid structure described 
in the previous Section
by numerical diagonalization of the 
system (\ref{unitary}). 

\begin{figure}
\epsfig{file=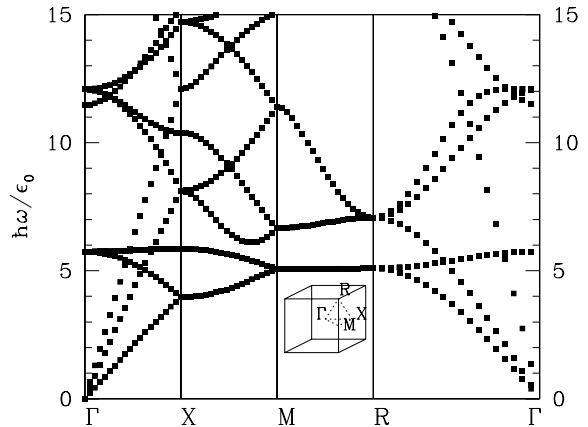,height=3.3 in,clip=}  
\caption{Calculated excitation 
dispersion along symmetry lines of the cubic lattice
1$^{st}$ Brillouin zone.
Energies are expressed in units of 
$\epsilon _0\equiv \hbar ^2/(MR_c^2)$.
}
\label{fig7}
\end{figure}

Our results are shown in Fig.\ref{fig7}
where $\omega ({\bf k})$
is plotted along symmetry lines in the 1$^{st}$ 
Brillouin Zone (BZ), as shown
in the inset of the figure.
At low k-values we find four modes: three correspond to the usual 
phonon bands (one longitudinal and two transverse 
modes), while the fourth, softer mode is 
associated with the broken gauge symmetry.
These modes can be seen, e.g., 
along the $\Gamma $-$X$ direction where
the first four small-q modes 
from lower to higher frequencies 
in Fig.(\ref{fig7}) are
the Bogoliubov mode, the (doubly degenerate) transverse 
phonon mode and the longitudinal phonon mode, respectively.

To unambiguously make the above assignments 
we computed \cite{saccani1}
the effective potential corresponding to the 
cluster solid structure,  
$V_{eff}({\bf r})=\int{ V(|{\bf r}-{\bf r} ^\prime|) |
\Phi ({\bf r} ^\prime)|^2d{\bf r} ^\prime }$
(see Eq.(\ref{schrod}))
and defined an effective force constant 
by fitting $V_{eff}$, close to its minima, with a quadratic curve. 
We then obtained the phonon frequencies by
solving the dynamical matrix for a harmonic crystal with such
force constants and an atomic mass equal 
to that of a cluster (i.e.
$\sim 10 \,M$). We find in this way two (degenerate)
transverse and one longitudinal modes, whose dispersion 
agree within 10-15\% at low q-values with 
the corresponding dispersion curves shown in Fig.(\ref{fig7}).

The local density and phase fluctuations for a given ${\bf k}$ value
and a given band $n$ are provided by \cite{macri}
\begin{eqnarray}
\Delta \rho _{n,\bf k} ({\bf r})=|u_{n,\bf k} ({\bf r})-
v_{n,\bf k} ({\bf r})|^2 \nonumber \\
\Delta \phi _{\bf k +\bf G} ({\bf r})=|u_{n,\bf k} 
({\bf r})+v_{n,\bf k} ({\bf r})|^2
\label{fluct}
\end{eqnarray}
By studying the spatial distribution of both 
density and phase fluctuations we find 
that, while the longitudinal phonon branch contributes mainly
to density fluctuations $\Delta \rho _{\bf k +\bf G} $,
the low-k Bogoliubov mode, which is also longitudinal in nature,
 contributes mainly to the phase fluctuations
$\Delta \phi _{\bf k +\bf G}$, being 
associated with the superfluid response.
These results parallel the similar findings in the 2-dimensional triangular
supersolid phase studied in Ref.\cite{macri}.

\begin{figure}
\epsfig{file=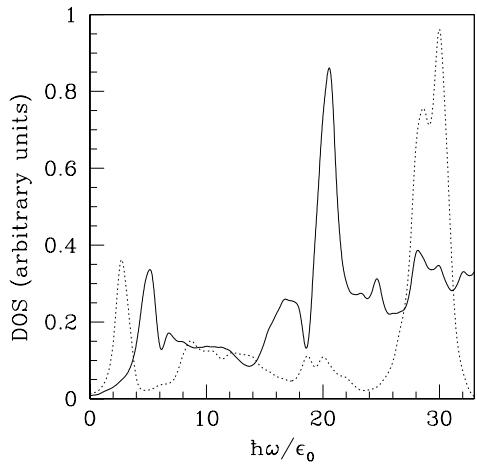,height=3.3 in,clip=}  
\caption{Calculated Density of States for two different values 
of $\Lambda$. Solid line: $\Lambda = 15.8$, dotted line: $\Lambda = 21$.}
\label{fig8}
\end{figure}

From the calculated 
excitation spectrum $\omega ({\bf k})$, one may obtain the density of states 
(DOS) as:
\beq
D(\omega )=\sum _{{\bf k}}\delta(\omega -\omega({\bf k}))
\label{dos}
\eeq
where the summation is restricted to the 1$^{st}$ BZ
of the crystal.
Fig.(\ref{fig8}) shows our calculated DOS. To compute it 
we have used a uniform 
mesh of 82 points within the irreducible part of the
1$^{st}$ BZ (shown with dotted lines in the
inset of Fig.(\ref{fig7})), corresponding to a sampling of 
4032 points in the whole 1$^{st}$ BZ. 

One can recognize two main peaks in the DOS: the one at low frequencies
(below $\sim 10\,\epsilon _0$), comes mainly from the zone-edge 
Bogoliubov soft mode,
while the other (above $\sim 20\,\epsilon _0$)) comes from the 
zone-edge longitudinal phonon modes.

The two curves (solid and dotted lines) shown in 
Fig.(\ref{fig8}) correspond to two different values of 
$\Lambda $, i.e. $\Lambda = 15.8$ and $\Lambda = 21$, respectively.
We notice that by increasing $\Lambda $ (e.g. by increasing density 
at constant $V_0$ and $R_c$) the phonon-like
excitation peak moves to higher frequency
because the particles get more localized around
the FCC lattice sites, making the crystal stiffer,
whereas the lower peak, associated with the 
Bogoliubov mode, shifts to lower frequencies, reflecting the 
loss of superfluid fraction. 

\section{\bf CONCLUSIONS} 

Quantum solid of clusters might be the 
prototypical system to realize
and study the supersolid phase of matter.
The system spontaneously breaks the translational invariance leading
to the appearance of a crystalline phase of individual 
superfluid droplets governed by a global macroscopic 
wavefunction. This system can exhibit supersolid behavior by decoupling 
a superfluid component from the crystalline structure.

Using mean-field approach, we have carried out a numerical study
of the structure and excitation spectrum of a soft-core model for
a supersolid.
We have computed the lowest energy structure: for values of the 
interaction parameter $\Lambda$  smaller than $\Lambda_{ss}=15.2$
the ground state is a uniform, superfluid phase, while for 
larger values it is a crystalline phase with FCC symmetry. 
Each unit cell contains clusters of atoms whose 
number increases with the system density,
for a given value of the interaction parameter.
 
Our results for the critical value of $\Lambda$ are in quite 
good agreement with exploratory PIGS Monte Carlo
results.
Due to the high computational cost of the {\it exact} Quantum Monte 
Carlo simulation, the present results are intended to be precursory of a
full ab-initio study of the 3D soft-core model that we leave for a future
work.

We have also found a range of values of the interaction parameter 
which would allow coexistence of the two (superfluid and supersolid) 
phases, and calculated the corresponding interface tension.

We have computed the supersolid excitation spectrum within the
Bogoliubov theory.
Besides the usual low-k phonon-like excitations
a new gapless, softer mode associated with the presence of a finite 
superfluid fraction appears, whose velocity decreases with increasing interaction parameter. 
The presence of this extra mode signals the breaking of gauge symmetry 
in the crystalline phase.
We have calculated how the fraction of superfluid density 
varies as a function of the interaction parameter, 
finding a discontinuous drop from 1 to $\simeq$0.4 at $\Lambda_{ss}$

In addition to the PIGS calculations mentioned above, 
the results of our calculations should provide a 
useful guide for experimental measurements.
As proposed in Ref.\cite{ruprecht} in fact, direct experimental 
measurements of the quasi-particle
excitation of a condensate system are possible
in principle by applying weak harmonic perturbation
to the trapping potential at some probe 
frequency at the end of the cooling
cycle and then probing the condensate shape by 
allowing the condensate to expand ballistically. 
By repeating the measurement with 
an incremented value of the probe frequency until the
maximum distortion of the condensate is found relative to the case
where no perturbation is applied, one determines the
resonance lines.
Another possible experimental way to probe superfluidity in cold gas
systems is represented by Bragg scattering \cite{mottl}.

\medskip
\begin{acknowledgments}
We acknowledge useful discussions with 
Tommaso Macri, Valerio Rizzi, Luca Salasnich and Milton W. Cole.
\end{acknowledgments}

\end{document}